\documentclass{emulateapj}

\newcommand{\dgr}{$^{\circ}~$}

\newcommand{\nhie}{$N_{\rm H \small{I}}$}

\newcommand{\hi}{H{\footnotesize I} }
\newcommand{\hie}{H{\footnotesize I}}

\begin{document}

\title{A Tidally-Stripped Stellar Component of the Magellanic Bridge}

\shorttitle{The SMC Line-of-Sight Depth}
\shortauthors{NIDEVER ET AL.}

\author{David L. Nidever\altaffilmark{1,2},
Antonela Monachesi\altaffilmark{1},
Eric F. Bell\altaffilmark{1},
Steven R. Majewski\altaffilmark{2}, \\
Ricardo R. Mu\~noz\altaffilmark{3},
Rachael L. Beaton\altaffilmark{2}
}
\altaffiltext{1}{Dept. of Astronomy, University of Michigan,
Ann Arbor, MI, 48104, USA (dnidever@umich.edu)}

\altaffiltext{2}{Dept. of Astronomy, University of Virginia,
Charlottesville, VA, 22904-4325, USA}

\altaffiltext{3}{Departamento de Astronom\'ia, Universidad de Chile, Casilla 36-D,
Santiago, Chile (rmunoz@das.uchile.cl)}

\begin{abstract}
%Deep, high-quality CTIO-4m+MOSAIC photometry of the Small Magellanic Cloud (SMC)
Deep photometry of the Small Magellanic Cloud (SMC)
stellar periphery ($R$=4\degr, 4.2 kpc) is used to study its line-of-sight depth with red clump (RC) stars.
The RC luminosity function is affected little by young ($\lesssim$1 Gyr) blue-loop stars in
these regions because their main-sequence counterparts are not observed in the color magnitude
diagrams.  The SMC's eastern side is found to have a large line-of-sight depth ($\sim$23 kpc)
while the western side has a much shallower depth ($\sim$10 kpc),
consistent with previous photographic plate photometry results.
We use a model SMC RC luminosity function to deconvolve the observed RC magnitudes
and construct the density function in distance for our fields.  Three of the eastern fields
show a distance bimodality with one component
at the ``systemic'' $\sim$67 kpc SMC distance and a second component at $\sim$55 kpc.
Our data are not reproduced well by the various extant Magellanic Cloud and Stream simulations. 
However, the models predict that the known \hi Magellanic Bridge (stretching
from the SMC eastward towards the LMC) has a decreasing distance with angle from the SMC and should be seen in
both the gaseous and stellar components.  From comparison with these models we conclude that
the most likely explanation for our newly identified $\sim$55 kpc stellar structure in the eastern SMC is a
stellar counterpart of the \hi Magellanic Bridge that was tidally stripped from the SMC $\sim$200 Myr ago
during a close encounter with the LMC.  This discovery has important implications for microlensing surveys
of the SMC.
%The existence of this sizable stellar structure in front
%of the main SMC stellar population will increase self-lensing and have important implications for microlensing
%surveys in this region of the SMC.
\end{abstract}


\section{Introduction}
\label{sec:intro}

The Large and Small Magellanic Clouds (LMC and SMC respectively) are the two
largest satellite galaxies of the Milky Way (MW) and, due to their proximity, offer the
best possibility for detailed study of dwarf galaxies, especially interacting dwarf irregulars.
One of the most striking features of the Magellanic system is the vast extent of
its \hi component -- including the 200\degr--long Magellanic Stream (MS) and Leading Arm \citep{Nidever10}
as well as the Magellanic Bridge \citep{Muller03}.  These gaseous structures are the result of past
interactions of the Magellanic Clouds (MCs) with each other and the MW galaxy
\citep{Murai80,GN96,Connors06,Ruzicka10,DB12,Besla10,Besla12,Besla13}.  The Magellanic Bridge is widely believed to
have been tidally stripped from the SMC by a recent close encounter with the LMC $\sim$200 Myr ago
\citep[e.g.,][]{MB07}.
%\citep[e.g.,][]{GN96,MB07,Ruzicka10,DB12}.

While the evidence of MC interactions is quite evident in \hi it is less obvious in the stellar structures of the Clouds.
The SMC, which is approximately ten times less massive than the LMC, is more likely to be affected by any past
interactions (and is the source of much of the MS \hi material in many models), and, therefore, may be the most obvious
place to search for {\em stellar} signs of tidal disturbances.

Two decades ago, a series of papers analyzing photographic plate photometry \citep{Hatz89} reaching to just below the SMC
horizontal branch (HB) performed the first detailed study of the SMC stellar periphery.  Using
red clump (RC) stars as standard candles in two fields, \citet{HH89} found the line-of-sight depth to be much larger in the
northeast than in the southwest.  Follow-up spectroscopy showed a correlation between distance and radial velocity (RV) in the
northeast RC stars \citep{Hatz93} similar to that seen by \citet{MFV86} in Cepheid variables closer to the center.
\citet{GH92} used the photographic plate photometry of all their SMC fields to trace HB stars to
$R$$\sim$5\dgr in all directions, uncovering a fairly symmetric structure but with a quick decline in density towards the west.
In contrast, the young main-sequence stars have a much more irregular shape extending towards the LMC into the
Magellanic Bridge region (see black contours in Fig.\ \ref{fig_map}).  \citet{GH91} found that the change of line-of-sight
depth with radius in the western SMC, increasing
with radius, was more consistent with a spheroidal than disk-like structure.  The notion of the stellar SMC having a spheroidal
structure is argued by \citet{Zaritsky00} and supported by
a recent spectroscopic study of $\sim$2000 red giant branch (RGB) stars in the central 4 kpc $\times$ 2 kpc
of the SMC that found no sign of rotation \citep{HZ06}.
In contrast, rotation is observed in the \hi component of the SMC \citep{Stani04}.
Many of the structural and kinematical features of the SMC are reproduced by the LMC-SMC-MW interaction simulations of
\citet{BK09} by using a spheroidal stellar distribution and extended gaseous disk.

More recent work on the SMC periphery has shown it to extend much further than previously thought. \citet{NG07}
detected intermediate-age and old stars in deep photometric data at $\sim$6\dgr south of the SMC center, while
\citet{DePropris10} found spectroscopically-confirmed RGB stars out to $R$$\sim$6\dgr in the eastern SMC.
\citet{Nidever11} used photometrically-selected RGB stars to trace the structure of the SMC periphery to
$R$$\sim$11\dgr  (in multiple directions) and showed that the SMC has a fairly azimuthally-symmetric structure.
Nidever et al.\ also found that the center of the outer SMC population ($R$$>$4\degr) is offset by $\sim$0.6\dgr (to the east)
from the center of the inner population ($R$$\lesssim$3\degr) and postulated that this is due to a perspective effect because,
on average, the stars to the eastern side are closer than the stars on the western side.
In addition, \citet{Bagheri13} used 2MASS and WISE catalogs to find evidence for some candidate older stars in the
region between the Magellanic Clouds, while \citet{Noel13} used deep photometry to find intermediate-age stars in a field at
$\sim$7\dgr from the SMC in the Magellanic Bridge.

Variable stars and stars clusters have been widely used to study the 3D structure of the inner SMC.  Several studies in
the 1980s found a large line-of-sight depth in the central SMC using young Cepheids \citep{MFV86,MFV88},
with \citet{CC86} finding hints of two arms in the southwest.
\citet{Haschke12} used multi-epoch OGLE-III \citep{ogle3} photometry to study the 3D distribution of
cepheids ($\sim$30--300 Myr old) and RR Lyrae ($\gtrsim$10 Gyr old) stars in the inner SMC ($R$$<$2\degr).
While both populations are found to have an extended scale height, the older stars have a fairly
homogeneous distribution whereas the young stars are in an inclined orientation that is closer 
in the northeast than in the southwest.
Finally, \citet{Crowl01} used populous SMC clusters to study the line-of-sight depth in the inner SMC
($R$$\lesssim$3 kpc) and found a $\pm$1 $\sigma$ depth between $\sim$6 and $\sim$12 kpc.

Even with all of these studies, the detailed 3D structure of the SMC stellar periphery is still not well understood
in large part due to the lack of high-quality, wide-area CCD photometry in this region of the southern
sky.  This is unfortunate because this knowledge would enable us not only to produce a better 3D map of the SMC but also provide
us with observational data of {\em collisionless} particles that would much better constrain the recent
($\sim$200 Myr) close encounter of the MCs with each other.  It is thought that during that encounter
the SMC might have passed right through the LMC disk \citep{Besla12} and produced the \hi Magellanic Bridge.

In this paper we use deep Washington $M$ and $T_2$ photometry from the MAgellanic Periphery Survey (MAPS)
to study the line-of-sight depth of core helium-burning RC stars in the SMC periphery.
We find a much larger line-of-sight depth ($\sim$23 kpc) in four eastern fields than in our four western fields ($\sim$10 kpc),
and a distance bimodality in three eastern fields with a newly identified component at $\sim$55 kpc
($\sim$12 kpc closer than the component at the ``systemic'', $\sim$67 kpc SMC distance)
that is most likely an intermediate-age/old stellar counterpart of the recently tidally-stripped \hi Magellanic Bridge.
In Section \ref{sec:data} we briefly describe the observations and main features of the color magnitude
diagrams (CMDs).  We argue in Section \ref{sec:nature} that the extended RC seen in many of our CMDs are
due to large line-of-sight depths and not population effects.  Density distributions as a function of distance
are derived in Section \ref{sec:distances} and compared to simulations in Section \ref{sec:models}.  Finally, a
discussion of the results and their implications are presented in Section \ref{sec:discussion} and a brief summary
in Section \ref{sec:summary}.

\section{Data}
\label{sec:data}

\begin{figure}[t]
\begin{center}
\includegraphics[angle=0,scale=0.40]{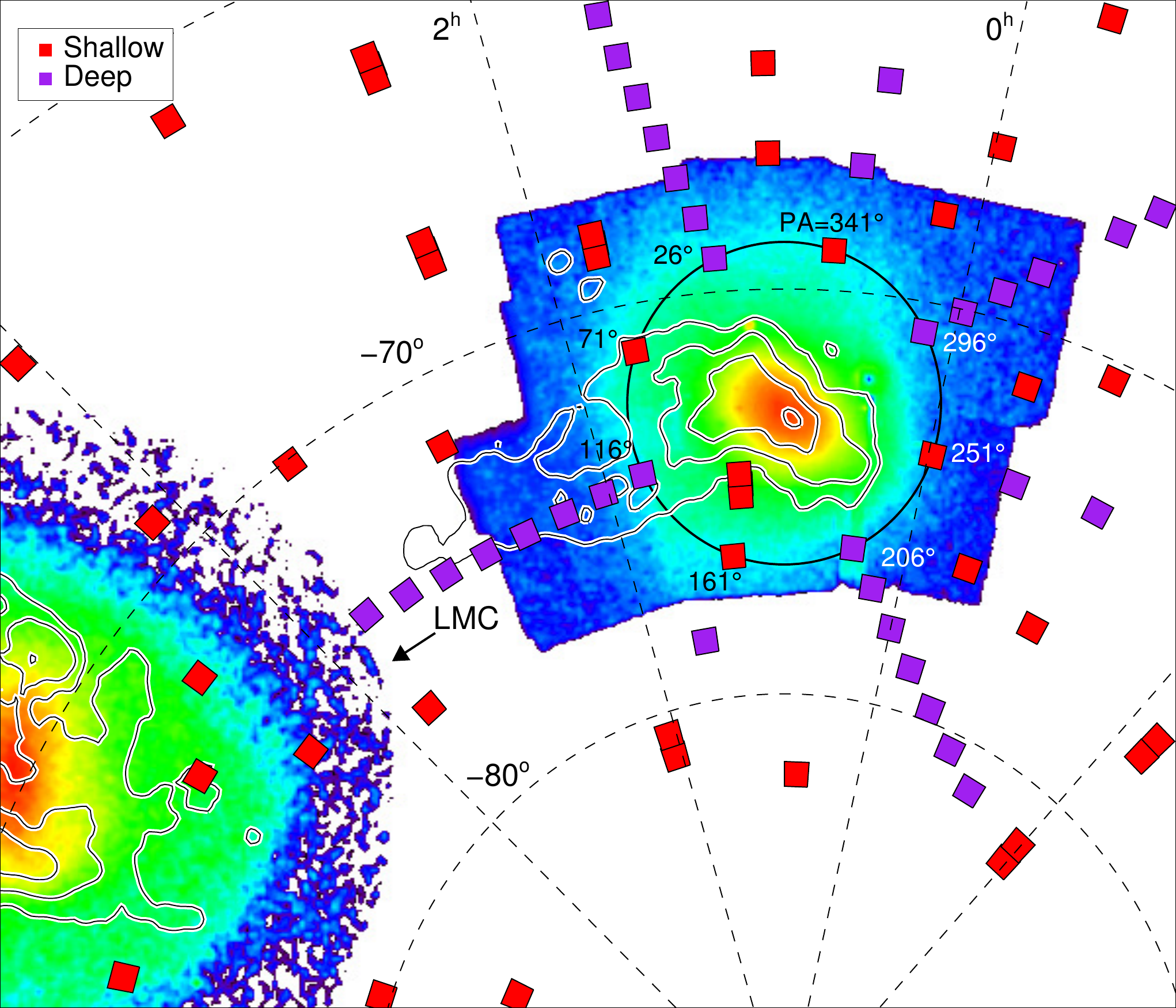}
\end{center}
\caption{Map of the SMC showing our CTIO-4m+MOSAIC fields as filled squares (red--shallow, purple--deep).
A circle at $R_{\rm SMC}$=4\dgr highlights the eight fields used in our analysis.
The colored image of the SMC shows the RGB starcounts using the combined Magellanic Clouds
Photometric Survey \citep{ZH02} and OGLE-III \citep{ogle3} photometry (for $R$$\lesssim$2\degr) and RC
starcounts from the photographic plate photometry of \citet{Hatz89} at larger radii.  The LMC is
shown in RGB starcounts selected from 2MASS \citep{Skrutskie06}.  The black contours indicate the \hi in
the Magellanic Clouds and Bridge from \citet{Bruens05} at levels of log(\nhie)=20.7, 21.1, 21.5 and 21.9.}
\label{fig_map}
\end{figure}

We use CTIO-4m+MOSAIC Washington $M$ and $T_2$ (equivalent to $I_{\rm C}$) photometry from the
MAgellanic Periphery Survey (MAPS) for our analysis.  The observations and data reduction are described
in \citet{Nidever11}.  Figure \ref{fig_map} shows the MAPS fields (filled squares) in the area of the SMC.
While we have ``deep'' photometry (to $M$$\sim$24) in fields extending to $R$$\approx$12\dgr the red
clump (RC) is most prominent in our $R$=4\dgr fields, which are the focus of the present study.

\begin{figure*}[t]
\begin{center}
\includegraphics[angle=0,scale=0.70]{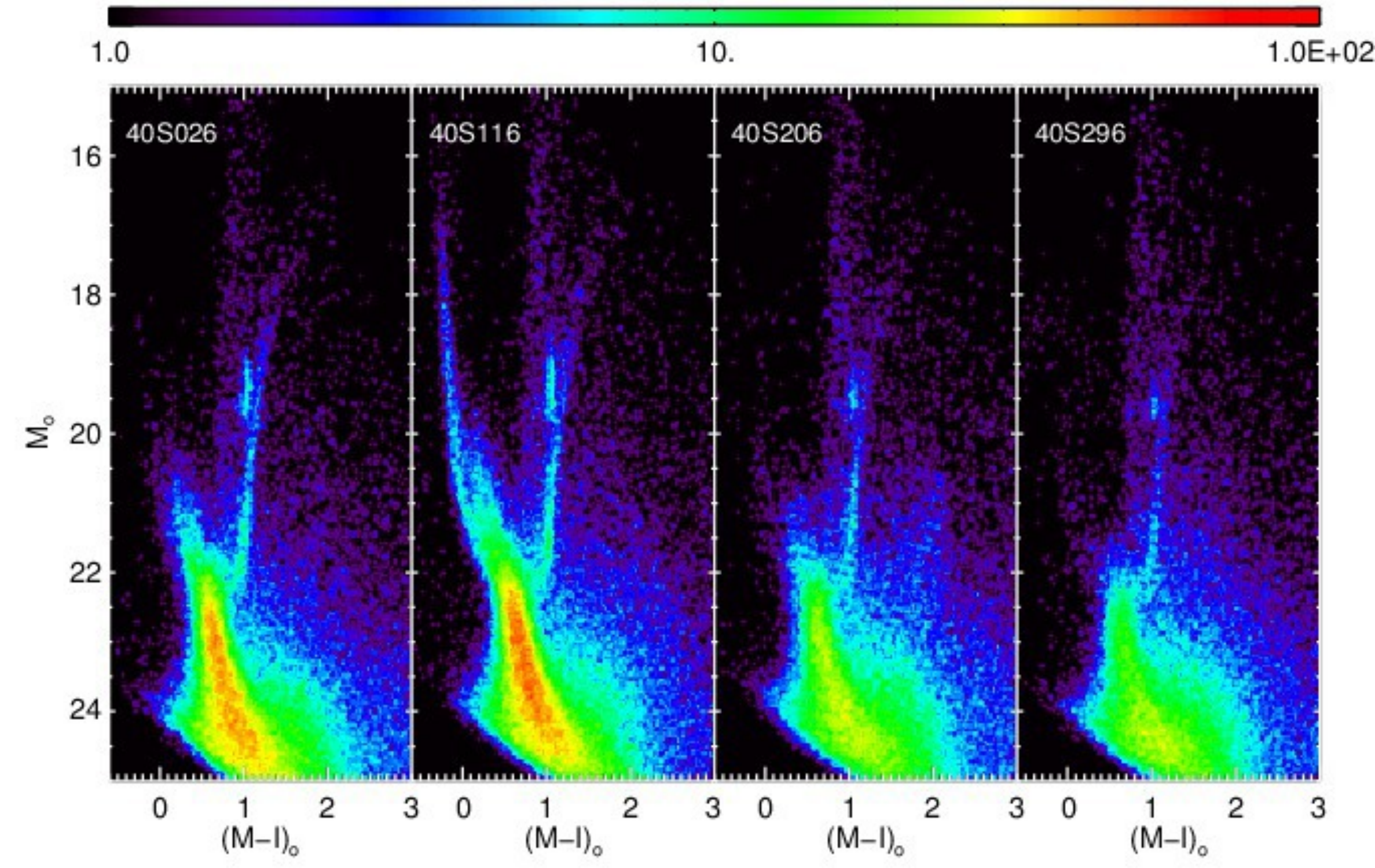}
\end{center}
\caption{Hess diagrams for the four deepest fields at $R$=4\dgr
showing the RGB, RC and main-sequence populations.  The extended RC can be seen in the two eastern
fields (40S026 and 40S116) while the two western fields display a much more compact RC.
The 40S116 field is in the \hi Magellanic Bridge where there is ongoing star formation and
young stars extending to bright magnitudes at $(M-T_2)_0$$\sim$$-$0.2 can be seen.}
\label{fig_cmdpanels}
\end{figure*}

\begin{figure*}[t]
\begin{center}
\includegraphics[angle=0,scale=0.65]{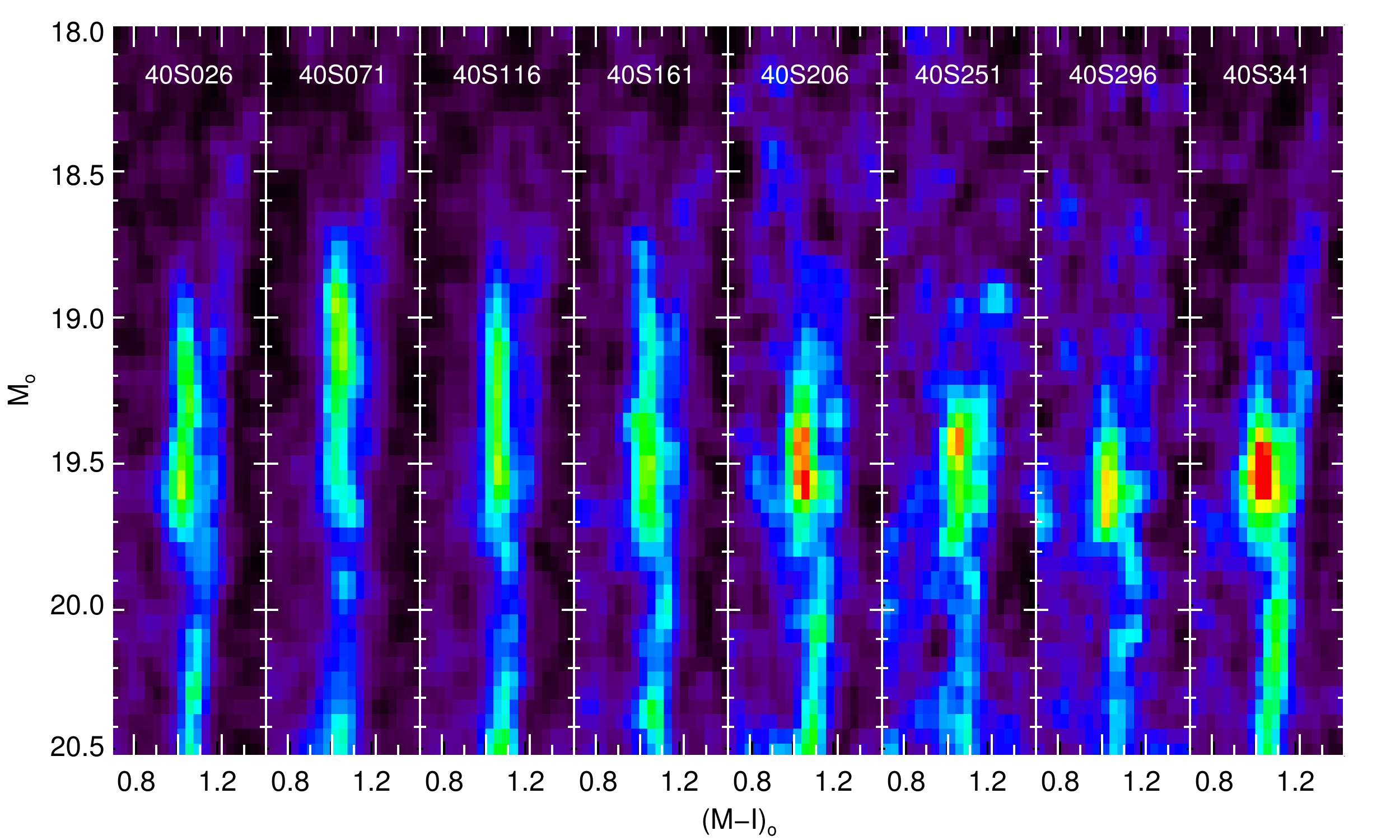}
\end{center}
\caption{Hess diagrams of the RC region for the
eight fields at R=4\degr (arranged from left to right with increasing position angle).  The Hess diagrams are scaled to the
same density of SMC stars to highlight the differences in RC morphology.  The four eastern fields (40S026--40S161)
show an extended RC that is not seen in the western fields (40S206--40S341).}
\label{fig_rcpanels}
\end{figure*}

Figure \ref{fig_cmdpanels} shows the full $M_0$ vs.\ $(M-T_2)_0$ Hess diagrams (dereddened with the Schlegel et al.\ 1998
extinction maps) of four, evenly spaced in azimuth, deep fields\footnote{The field names are
constructured from the field's radius and position angle (east of north), i.e. 40S026 is R=4.0\dgr and PA=26\degr.} at
$R$=4\degr.  The SMC RGB, RC, and main-sequence stars are clearly visible.
Hess diagrams of the RC region for all eight $R$=4\dgr fields are shown in Figure \ref{fig_rcpanels}.
The RC morphology clearly varies substantially from field to field.  The eastern fields (PA=26--161\degr) show a much
larger RC extent in magnitude than the western fields (PA=206--341\degr).  It was previously noted by \citet{HH89}
that the horizontal branch stars in the northeast of the SMC periphery have a larger line-of-sight depth
(and extend to brighter magnitudes) than in the southwest.

\section{The Nature of the Extended Red Clump Luminosity Function}
\label{sec:nature}

An elongated distribution of the red clump (to the bright end) can be caused by reasons other than a
large line-of-sight depth.  The evolution of intermediate-mass
He-core burning stars moves them along loops in the H-R diagram at nearly constant luminosity
(``blue loop'' stars).  Their age at a given magnitude depends on metallicity but blue loop stars are generally
quite young (a few hundred Myr to 1 Gyr; \citealt[][and references therein]{Sweigart87, XuLi04}) and theoretical
models predict that their CMD location depends strongly on metallicity \citep[e.g.,][]{Girardi00}.
Many stars pile up on the red end of the blue loops (BL) and define
a nearly vertical feature that stretches to brighter magnitudes than the RC (often spanning
$\sim$2-3 mag) and trending to the blue \citep[as seen in Fig.\ 3 of][]{Gallart98}.  This
feature of stellar evolution can be easily confused with an extended red clump and was the
primary point of contention regarding the nature of the ``vertical red clump'' of the LMC identified by
\citet{Zaritsky97}; these authors interpreted the feature as an intervening stellar population but this was
subsequently called into question by \citet{Beaulieu98}.
Given this precedent, we have taken steps to rule out BL stars as the cause for the extended feature observed
in the CMDs of the eastern SMC fields.

We have computed a synthetic CMD that reproduces the overall features of the stellar
populations of the SMC in the CMD (see Fig.\ \ref{fig_cmdpanels}) to compare
its RC and BL distributions with the elongated feature observed at the
RC level.  We have assumed a constant star
formation rate (SFR) from 0.7 to 12 Gyr and metallicities of [Fe/H]=$-1$ dex
([Fe/H]=$-1.5$ dex) for stars younger (older) than 6 Gyr
\footnote{Since these are the estimated ages and metallicities of the fields under study, as suggested in
\citet{Nidever11} by eye-fitting isochrones}.
The model CMD was computed using the IAC-STAR code
\citep{Aparicio04} adopting the BaSTI stellar library \citep{Pietrinferni04}, 
a Reimers mass loss efficiency parameter value of $\eta$=0.2,
and a \citet{Kroupa02} initial mass function (IMF) from 0.1 to $100M_{\odot}$.
The magnitudes of the model CMD are
expressed in the Johnson-Cousins photometric system; in particular we have chosen the bolometric correction library from
\citet{Girardi02}.  The magnitudes were transformed into the Washington $M$ and $T_2$ photometric system using the
transformation equations from \citet{Majewski00}.
We assumed a distance modulus of 18.9 for the SMC and
did not correct the model CMD due to observational effects (incompleteness and photometric errors).
Given that we expect a near 100\% completeness and a $\sim$2\% photometric accuracy ($\sim$0.3\% internal precision)
at the RC level, this correction should have little impact on the model CMD for our purposes.

\begin{figure*}[t]
\begin{center}
$\begin{array}{cc}
\includegraphics[angle=0,scale=0.35]{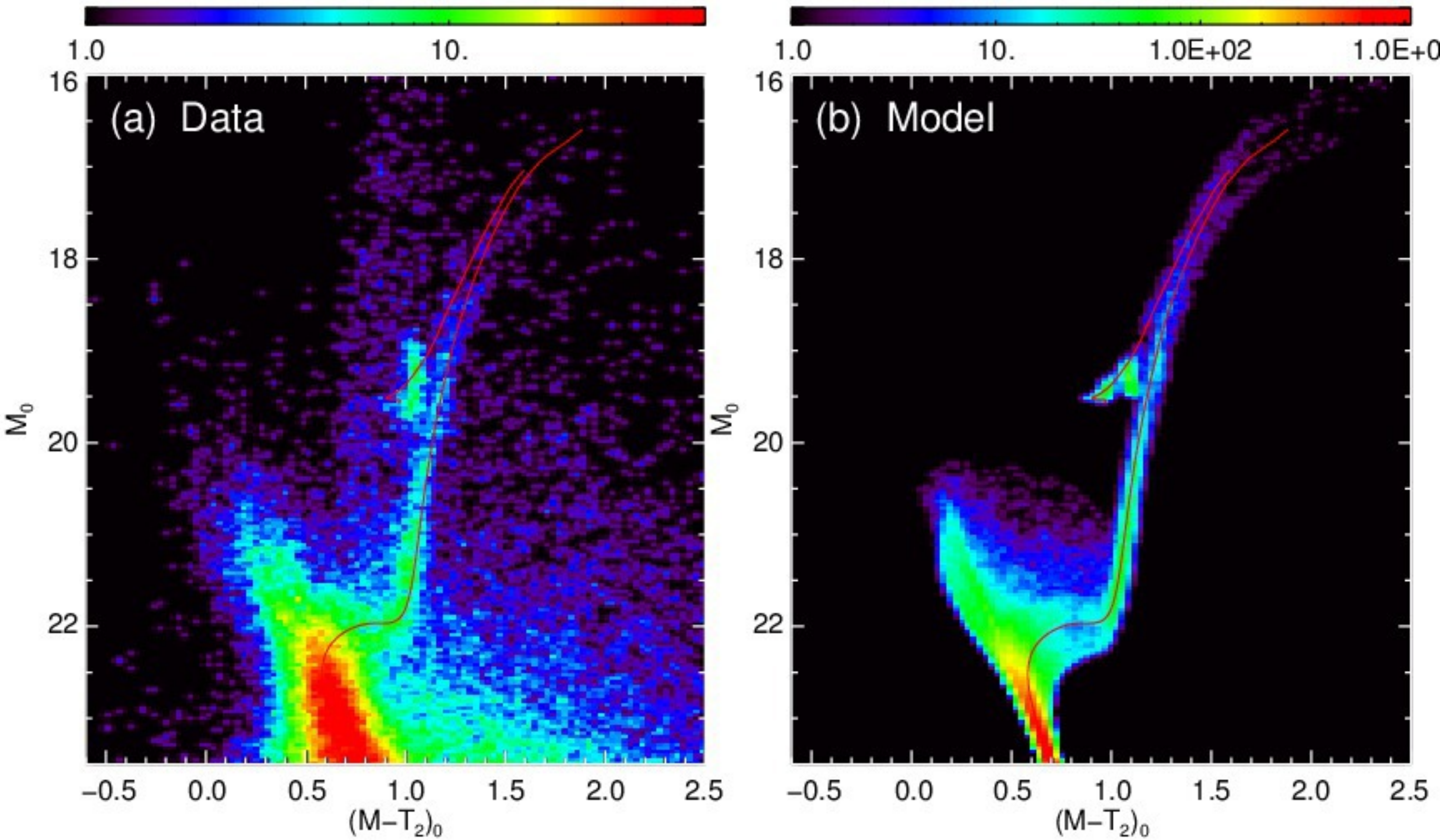} &
\includegraphics[angle=0,scale=0.35]{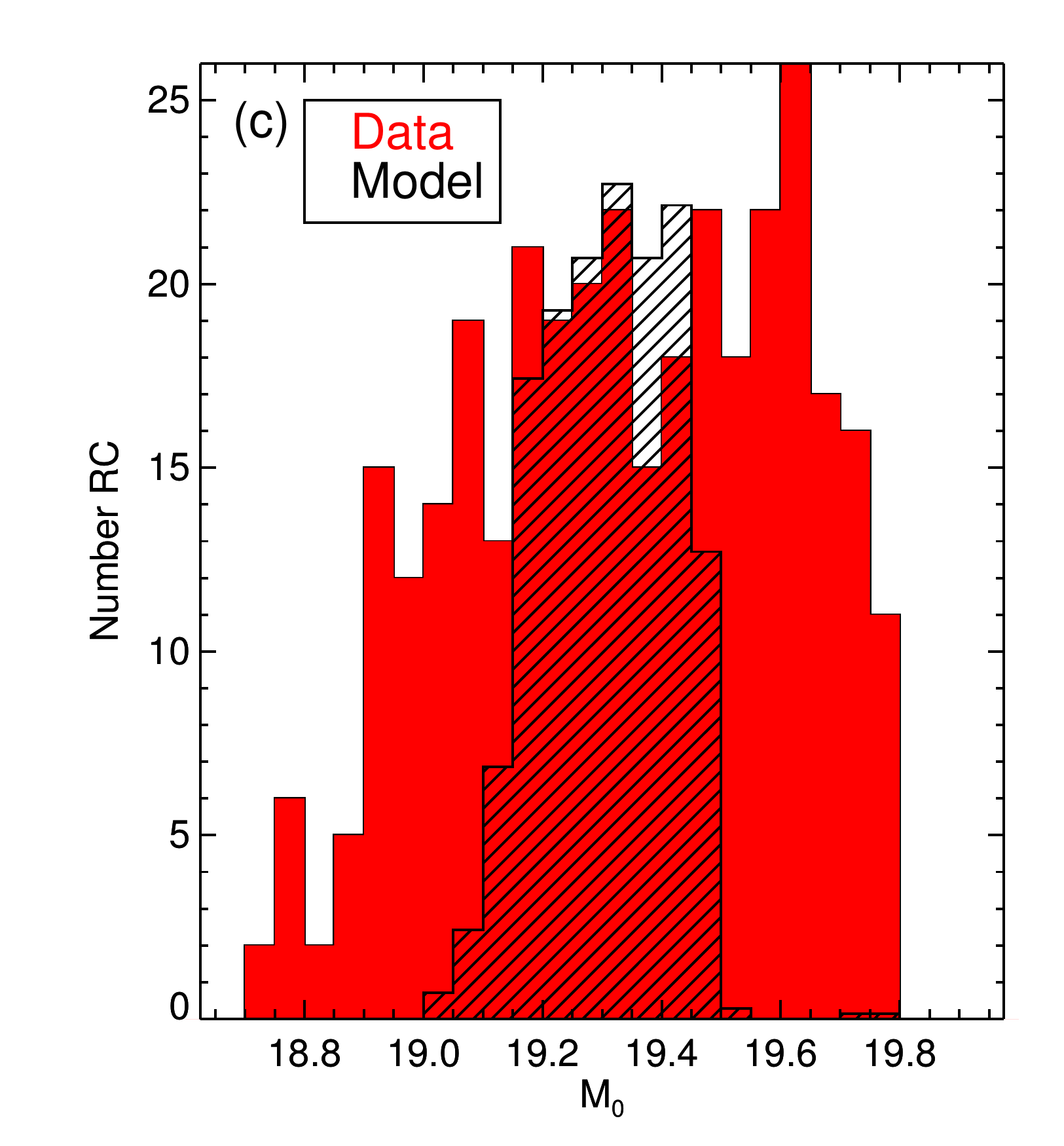}
\end{array}$
\end{center}
\caption{{\em (a)} Hess diagram of the 40S026 field in the eastern portion
of the SMC. {\em (b)} Simulated CMD for the 40S026 field with ages of 1.4 to 12 Gyr.
The red lines are [Fe/H]=$-$1.488 age=8.0 Gyr BaSTI isochrones \citep{Pietrinferni04} at 60 kpc.
{\em (c)} Red clump luminosity function for the data (red) and simulation (black).
The observed luminosity function is much broader than the model suggesting a large
line-of-sight depth.}
\label{fig_simcmd_east}
\end{figure*}

\begin{figure*}[t]
\begin{center}
$\begin{array}{cc}
\includegraphics[angle=0,scale=0.35]{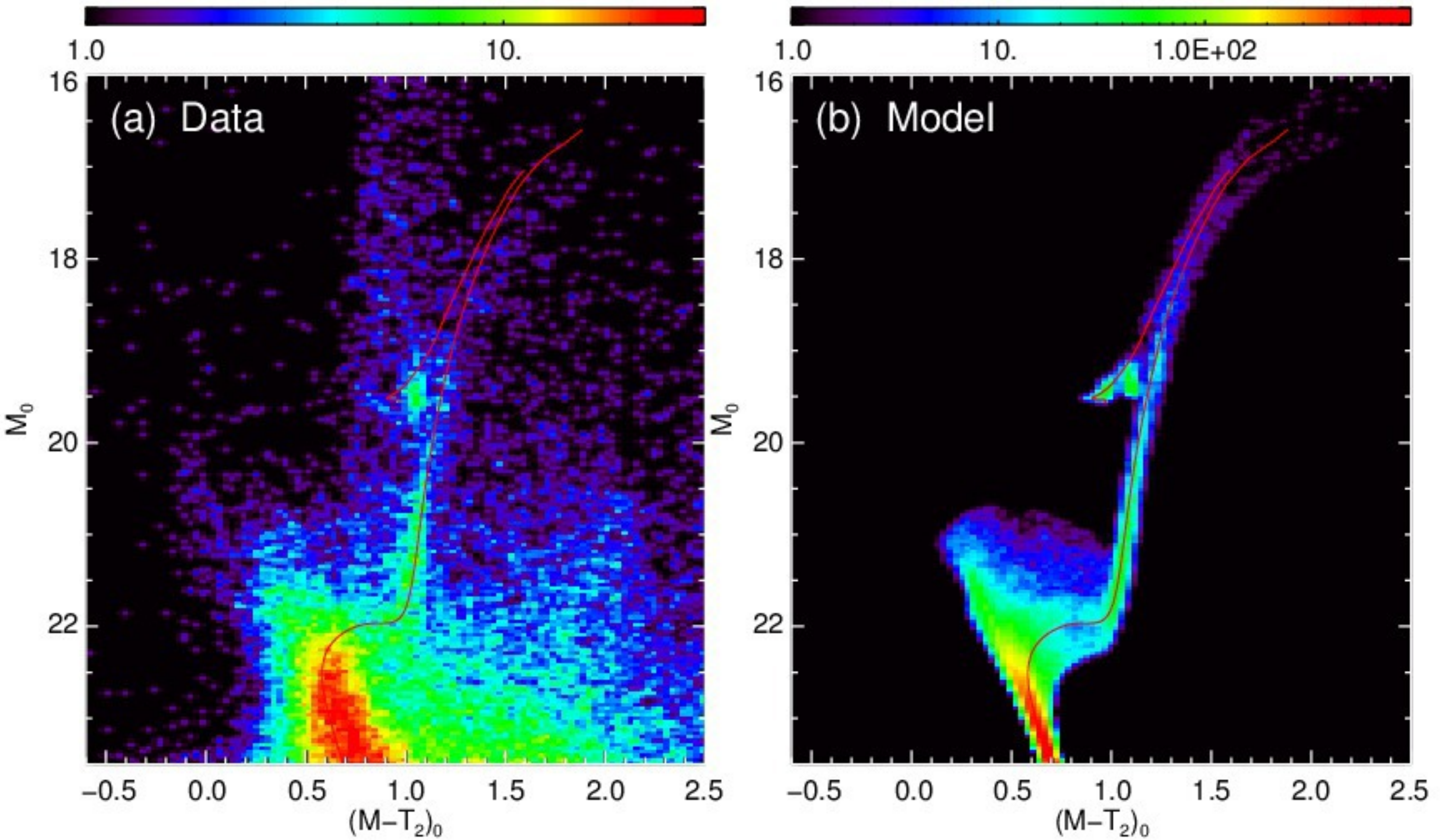} &
\includegraphics[angle=0,scale=0.35]{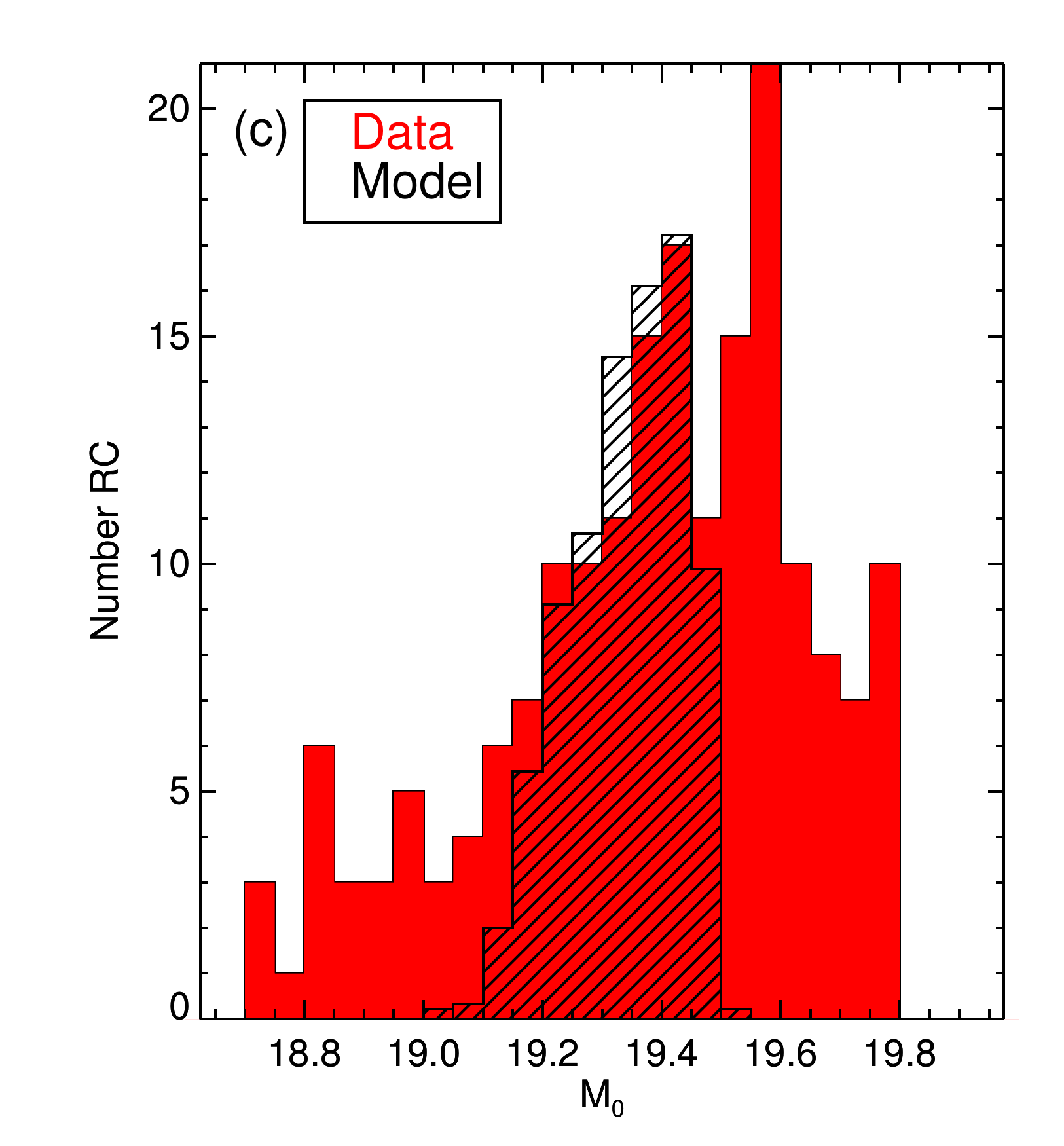}
\end{array}$
\end{center}
\caption{Same as Figure \ref{fig_simcmd_east} but for the 40S206 field with model ages of 2.0 to 12 Gyr.
While not as wide as the RC luminosity function of 40S026, the observed luminosity
function of 40S206 is wider than the model and indicates some depth is needed to
explain the observations.}
\label{fig_simcmd_west}
\end{figure*}

\begin{figure*}[t]
\begin{center}
\includegraphics[angle=0,scale=0.65]{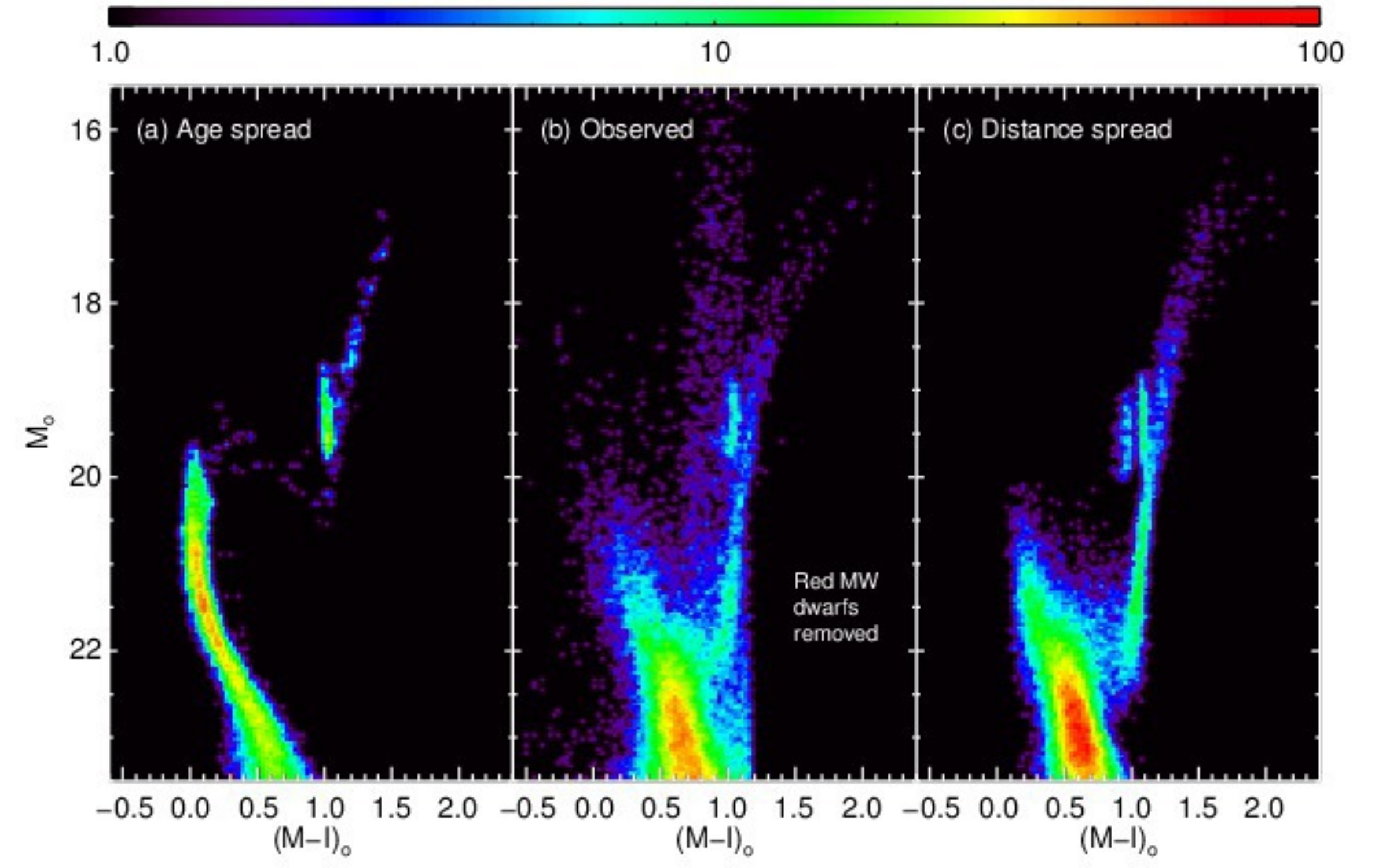}
\end{center}
\caption{{\em (a)} A simulated Hess diagram for 40S026 using the synthetic stellar populations and 
{\em age/SFR spread only} to reproduce the morphology of the RC region (with added photometric noise
using the formal observational uncertainties).
To achieve this, nearly all of the young and old populations are removed to leave many stars with 0.7--1.2 Gyr.
This is clearly not a good representation of the entire observed CMD. {\em (b)} Observed Hess diagram of
40S026 with red MW dwarfs removed, using the $(M-T_2,M-$DDO51$)_0$ diagram \citep{Majewski00}, to reveal the structure of the
upper RGB. {\em (c)} A simulated Hess diagram for 40S026 using synthetic stellar populations ($\geq$1.4 Gyr)
convolved with a distance spread (as derived in \S \ref{sec:distances} for this field) and added photometric
noise (as in panel {\em a}).  This is a good representation of the observed data.}
\label{fig_cmdconvolve}
\end{figure*}

Figures \ref{fig_simcmd_east} and \ref{fig_simcmd_west} show the observed (a) and model (b)
CMDs for the 40S026 (eastern SMC) and 40S206 (western SMC) fields, respectively. The model in Figure
\ref{fig_simcmd_east}b has ages ranging from 1.4 Gyr to 12 Gyr whereas the model in Figure \ref{fig_simcmd_west}b
has ages ranging from 2.0 Gyr to 12 Gyr. These age ranges were chosen so that the position and
characteristics of the main features (main sequence, subgiant branch,
tip of the RGB) are well reproduced by the models. In Figure \ref{fig_simcmd_east},
however, there is a clear difference in the RC morphology between the
model and the data. We find that the RC is much less elongated in
magnitude in the model than in the CMD of the observed eastern
field. This discrepancy cannot be accounted for by stellar
evolutionary features such as BL stars; the model predicts that there
should be almost no BL stars, given that its youngest population has
an age of 1.4 Gyr. A CMD model with stars as young as 0.7 Gyr predicts
more BL stars, but also shows evidence of a brighter and more populous
young main-sequence, which is not seen in the data.
Moreover, even if the young main-sequence
from the model CMD were to agree with the data, the density of the model BL
stars would still be significantly lower than that of the RC stars.
Figure \ref{fig_cmdconvolve}a is an attempt to reproduce the morphology of the RC region of the CMD (seen in
panel b) using a spread in age and star formation rate {\em only} (which requires most of the young and old stars to
be removed).  The resulting simulated Hess diagram
is {\em not} a good representation of the observed data, particularly on the main sequence.  On the
other hand, the simulated Hess diagram with a {\em distance}
spread (convolved with the distance function for this field found in \S \ref{sec:distances}) in
Figure \ref{fig_cmdconvolve}c is a good representation.
Thus, \emph{a BL population cannot explain the elongated RC that we observe
in the eastern field.} On the other hand, Figure \ref{fig_simcmd_west} shows that the
western field and model RC morphologies agree fairly well.  Figures \ref{fig_simcmd_east}c and
\ref{fig_simcmd_west}c show the RC luminosity
function both for the data and the model. The RC stars were isolated as explained in the
next section. Note the much wider RC distribution in the eastern field
when compared with the model.

\section{Distances}
\label{sec:distances}

\begin{figure*}[ht!]
\begin{center}
\includegraphics[angle=0,scale=0.77]{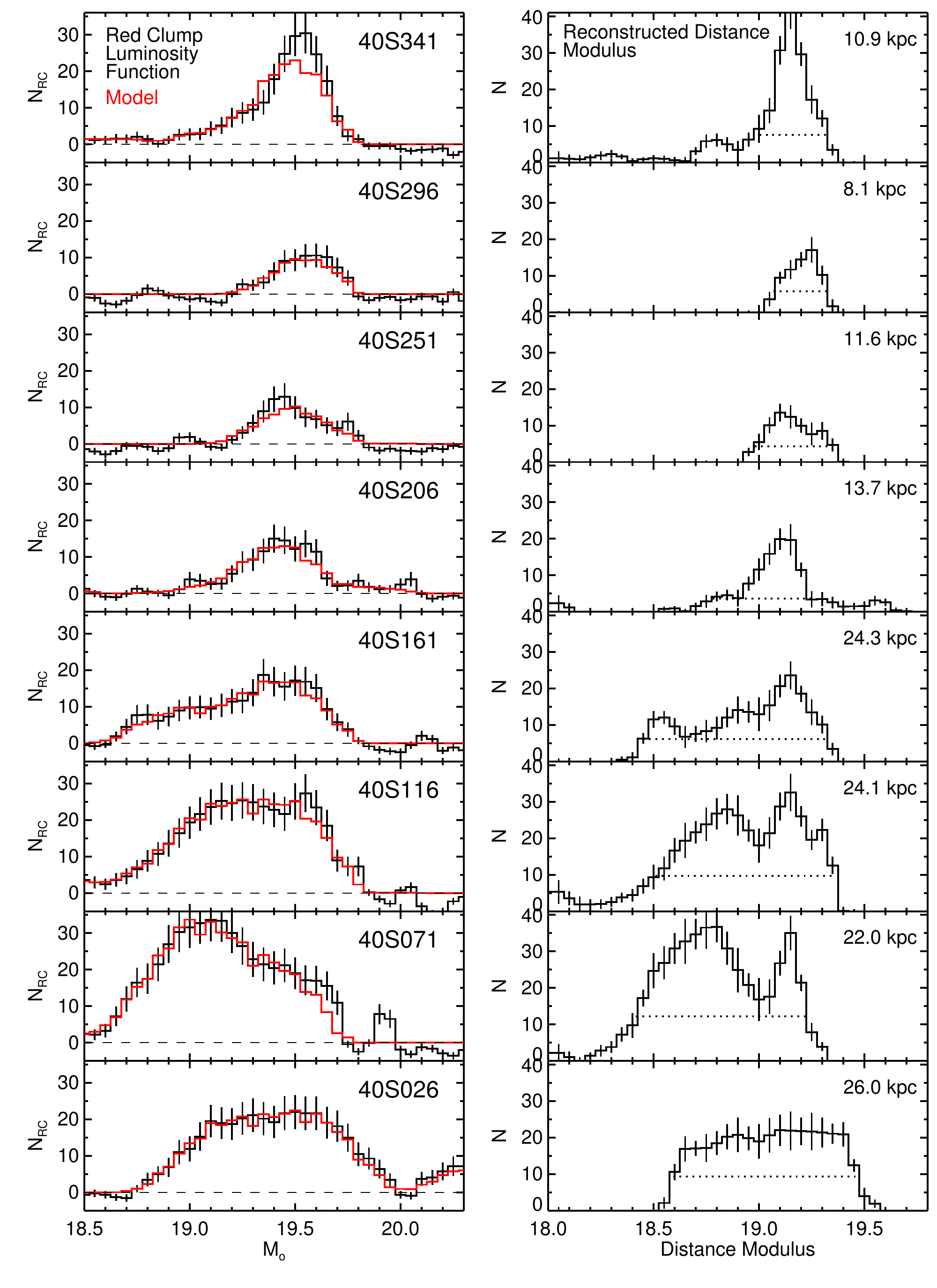}
\end{center}
\caption{(left) Red clump luminosity functions for the eight R=4\dgr fields (with Poisson errors).  Some residual RGB stars are
visible at faint magnitudes in 40S026 and 40S071.  The best-fitting model from the iterative procedure (see text for details)
is overplotted in red.  Overall the models fit well except for 40S341 which might require a narrower intrinsic RC function
to fit the data. (right) Reconstructed density function with distance modulus.  The errorbars show internal uncertainties found with
a Monte Carlo simulation. Spans of the distribution used to calculate the depth (in upper right-hand corner) are shown as
dotted lines.  Three of the eastern fields (40S071--40S161) show
bimodal distance distributions that have been enhanced through the reconstruction process.}
\label{fig_rclumpanels}
\end{figure*}

To study the distance distributions, we isolate the RC stars in the CMD.  Because the RC and RGB are
very close in the CMD and sometimes overlap we decided to model the RGB distributions to 
produce a ``clean'' RC sample.
A Hess density map of stars was created for each field in ($M-I$)$_0$ and $M_0$ in bins of 0.02 and 0.05 mag
respectively.  These were then smoothed with a Gaussian kernel having FWHM=2 bins.  Each row, at a given
magnitude, was modeled with a double-Gaussian for the RC and RGB over the magnitude range 18.5$<$$M_0$$<$20.5.  After
the first iteration over the magnitude range, robust linear fits with magnitude were performed to the
RGB Gaussian heights, centers
and widths.  On the second iteration, the RGB Gaussian components were constrained to lie close to the linear
fit values (especially the centers).  The final RGB model was then subtracted from the density map.  The
final image was summed over RC colors, 0.94$<$($M-I$)$_0$$<$1.12, and a similarly sized region blueward of the RC
(for MW foreground) was summed and subtracted.  The final RC luminosity functions are shown in Figure \ref{fig_rclumpanels}a
(with Poisson errors).  In some fields there is a residual RGB signal left at faint magnitudes
($M_0$$\gtrsim$20 or RC $d$$\gtrsim$80 kpc) in the subtracted density image (which can be seen in some
of the luminosity functions).  However, no obvious signs of an extension of the RC are seen at these magnitudes
in the CMD and so the RC luminosity functions for $M_0$$\gtrsim$20 are ignored for the rest of the distance analysis.

\begin{figure}[ht!]
\begin{center}
\includegraphics[angle=0,scale=0.45]{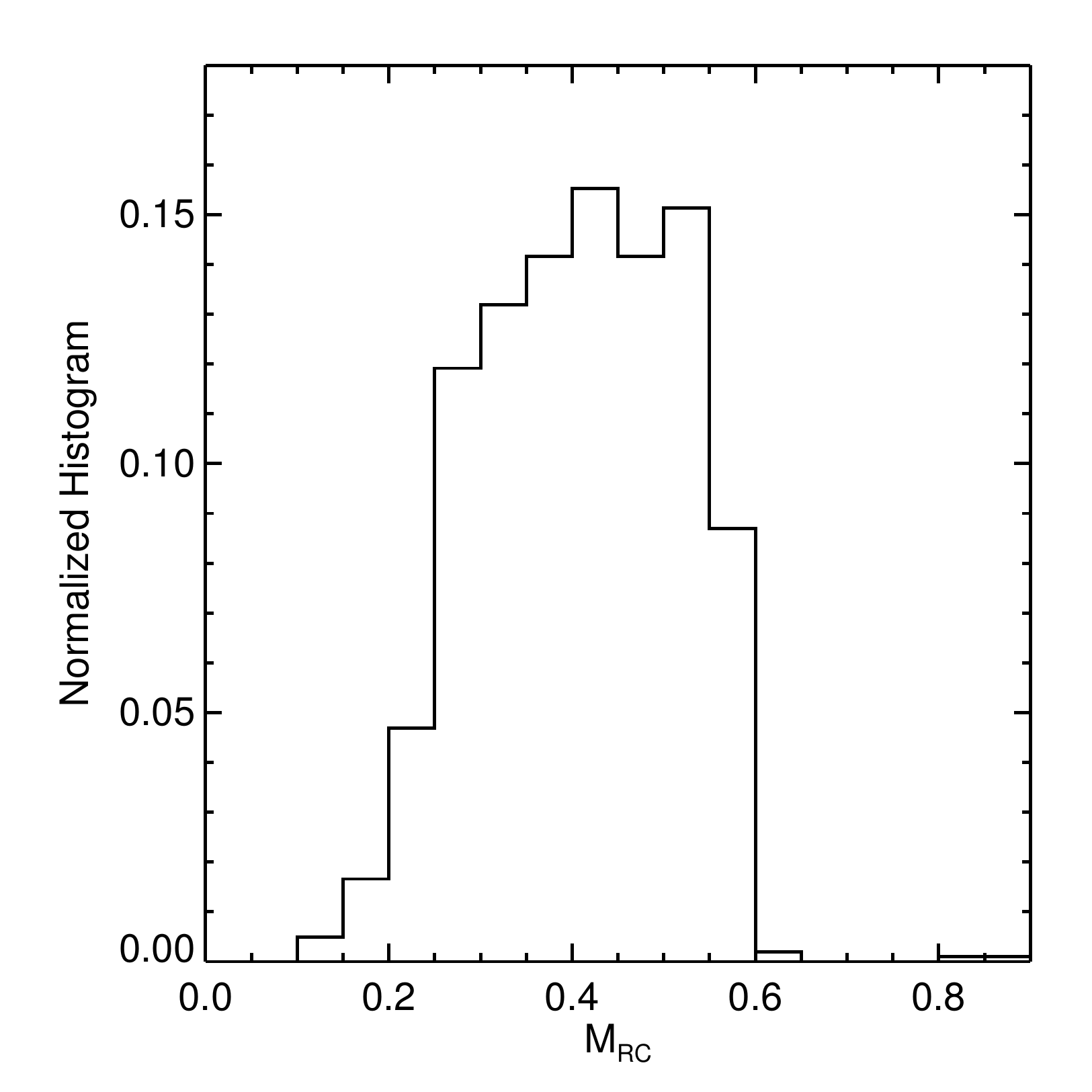}
\end{center}
\caption{The red clump absolute magnitude luminosity function using the SMC simulated CMD.  Ages from 1.4 to 12 to Gyr are
included.  The mean magnitude from a Gaussian fit is $M_{\rm RC}$=$+0.41$.}
\label{fig_rcabsmag}
\end{figure}

\begin{figure*}[t]
\begin{center}
\includegraphics[angle=0,scale=0.70]{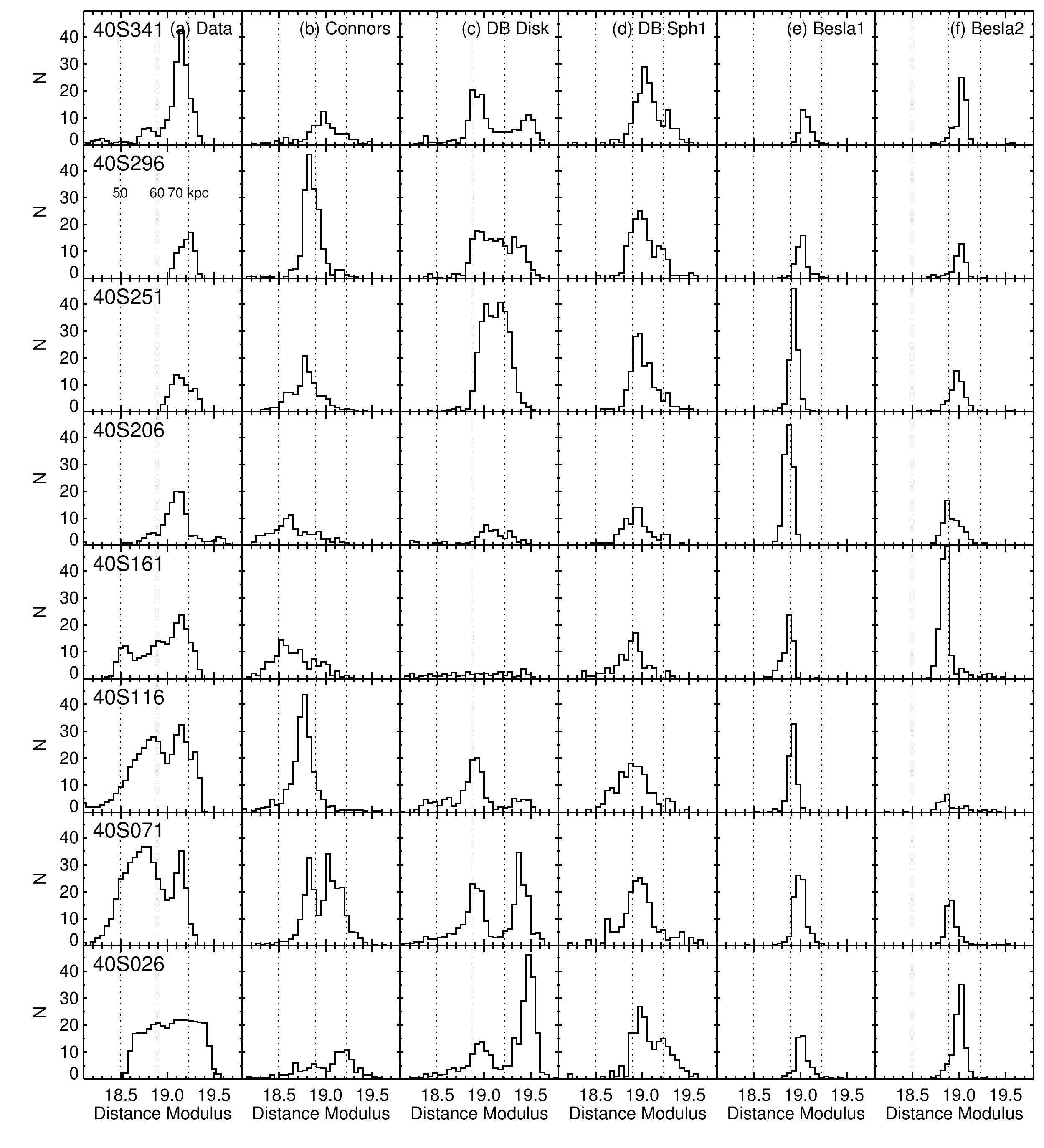}
\end{center}
\caption{Density function with distance modulus for our red clump data and various models:
{\em (a)} data, {\em (b)} \citet{Connors06} model (scaled by 1/2.5), {\em (c)} \citet{DB12} model disk component
(scaled by 1/4), {\em (d)} \citet{DB12} model spheroid1 component,  {\em (e)} \citet{Besla12} model1 (scaled by 1/3.5),
and {\em (e)} \citet{Besla12} model2 (scaled by 1/5).  The field names are given in the upper left hand
corner of column {\em a}.  Vertical dotted lines indicate 50, 60 and 70 kpc.}
\label{fig_models}
\end{figure*}

Next, we use the model SMC CMD to construct an absolute RC luminosity function
(isolating RC stars with 1.025$<$$M-I$$<$1.12 and 18.7$<$$M$$<$19.8)
and reconstruct the density function in distance (or rather distance modulus) for each field's
observed RC luminosity function.  A technique somewhat similar to the ``annealing''
method (which iteratively finds the state of maximum entropy) was used to derive the density function.  A discrete density function
in distance modulus (in 0.05 mag steps) is convolved by the absolute luminosity function (Fig.\ \ref{fig_rcabsmag})
to produce a model RC luminosity function that can be compared to the observed data and used to calculate $\chi^2$ (using
Poisson errors).  The reconstructed density function in distance is then found by an iterative approach.  The density function
is initialized with the observed RC luminosity function shifted by $-0.41$ mag (the mean
SMC model RC magnitude is $M_{\rm RC}$=+0.41 mag).  Each distance bin is then successively stepped through and its density
varied until the best-fitting $\chi^2$ value is found between the RC luminosity data and model.  After stepping through all
the distance bins the density function is smoothed with a FWHM=1.0 bin Gaussian kernel to smooth small-scale fluctuations.
This process is
then iterated many times until convergence (normally after $\sim$10 iterations) is achieved.  To ascertain internal uncertainties
in the derived density functions with distance, we performed a simple Monte Carlo simulation for each field.  Poisson noise was
added to the RC luminosity function and the iterative procedure performed.  This was repeated 50 times and the standard
deviation for each distance modulus bin (over the 50 mocks) was calculated and used as the internal uncertainty.  The final density
functions in distance modulus (and the uncertainties) can be seen in the right panels of Figure \ref{fig_rclumpanels} and the
best-fitting models (red) in the left panels.  To estimate the depth of a field we used the span of the curve at a density
level that ``bisects'' the distribution (i.e., half the area under the curve falls below this line and half above).  The spans
are shown as dotted lines in right panels of Figure \ref{fig_rclumpanels} and the corresponding depths are indicated in the 
upper right-hand corner.

The models are not perfect matches to the data and any small-scale structure in the density functions in distance should not
be taken to represent real structures.  However, we can use the density functions to discern broad features.  The eastern fields
(40S026--40S161) show large line-of-sight depths ($\sim$23 kpc) over a position angle range of 135\degr, while the western
fields have much shallower depths of $\sim$10 kpc.  Furthermore, three of the eastern
fields (40S071, 40S116 and 40S161) show evidence for a distance bimodality (with one component at $\sim$55 kpc and the second
at $\sim$63 kpc) and the fourth eastern field (40S026) has a broadened distribution, and is potentially consistent with the
trend seen in the other three fields.

\section{Comparison to Models}
\label{sec:models}

To help understand the nature of the large depth and bimodality in the eastern fields we compare our density
functions to various simulations of the Magellanic Clouds and Stream: \citet{Connors06}, \citet[][hereafter DB12]{DB12}, and
\citet[][hereafter B12]{Besla12}.  For each simulation, particles were selected at our field locations relative to the center of the
SMC in the simulation (which was sometimes shifted slightly from the observed center).  There were often not enough model
particles within the 0.36 deg$^2$ area of our field sizes to make useable distance histograms.
Therefore, a matching radius of 0.5\dgr was used for the Connors and DB12 models,
and 0.7\dgr for the DB12 models.
Figure \ref{fig_models} shows (a) our density functions, (b) the model of \citet{Connors06}, 
(c) the DB12 disk model, which these authors suggest primarily represents the \hi component
of the SMC, (d) the DB12 spheroid1 model, which they suggest represents a spheroidal-shaped stellar
component of the SMC, and (e) the B12 model1 and (f) model2 (both with stars older than 1 Gyr).  It is quite immediately
clear that none of the models adequately reproduce the shape and line-of-sight depth of the observed fields, although this is
not entirely surprising given that these models were optimized to reproduce the gaseous Magellanic Stream.
The DB12 disk model does show a bimodality
in some of the eastern fields caused by the main SMC body at $\sim$60 kpc and the ``counter-bridge'' (see section 3.5 of DB12)
at $\sim$80 kpc.  In contrast, however, the two components in the data appear at $\sim$67 kpc (likely the main SMC body) and at
$\sim$55 kpc (a newly-found stellar component) with very few stars beyond $\sim$70 kpc (except in 40S026).
Therefore, it is unlikely that the observed bimodality is related to the DB12 counter-bridge
(which our data effectively rule out as a stellar feature at these positions, 
though it could still exist at smaller radii).  We note that the counter-bridge is not very prominent
in the spheroid1 model (the model most likely to represent the stars) and it therefore might effectively be an \hie--only feature
(similar to the Magellanic Stream).

\begin{figure}[t]
\begin{center}
\includegraphics[angle=0,scale=0.57]{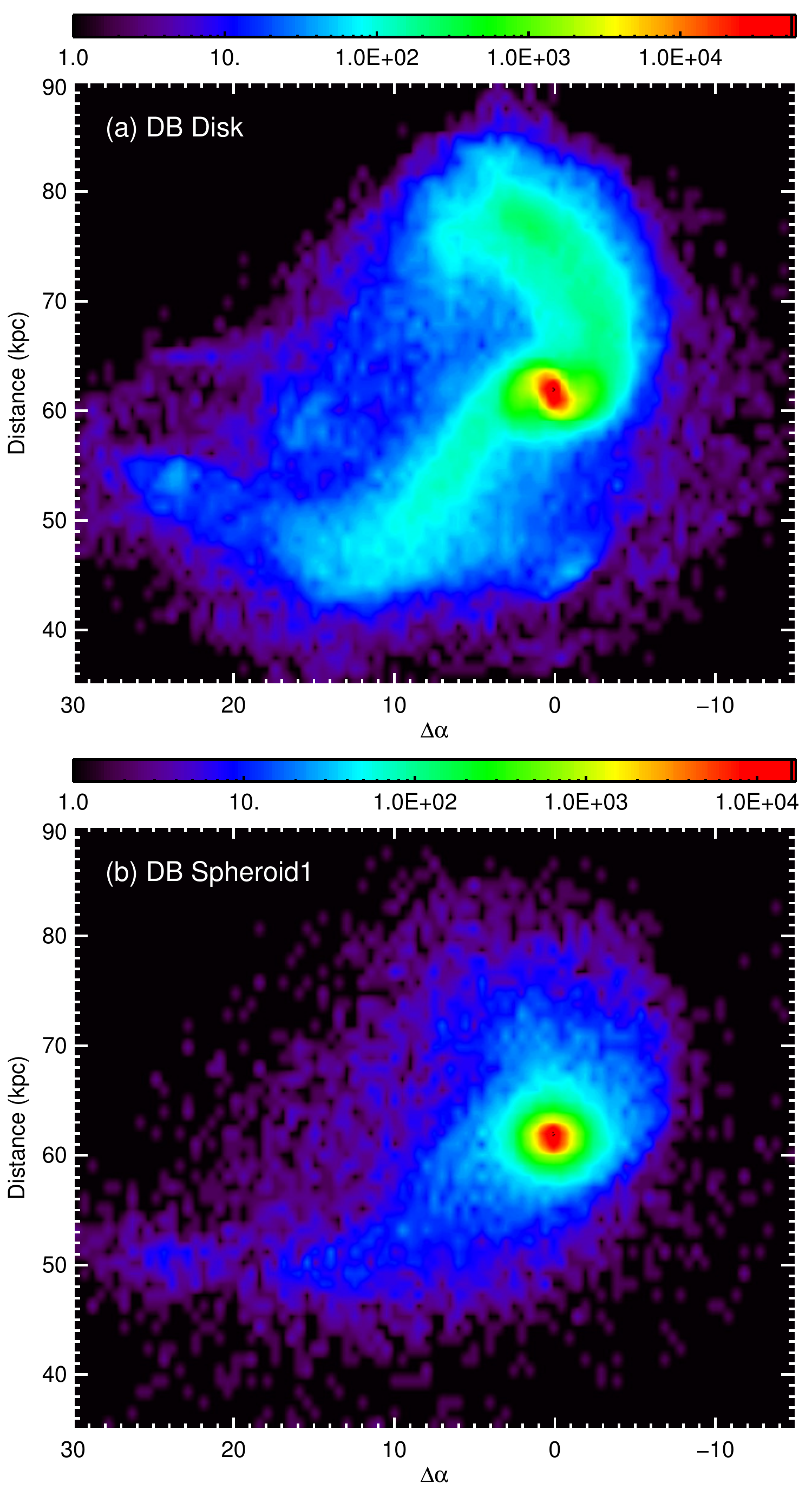}
\end{center}
\caption{Distance--$\Delta\alpha$ diagrams for the \citet{DB12} disk (a) and spheroid1 (b) models
(for particles with $|\Delta\delta|$$<$10\degr).
Both models show the bridge extending to the east and closer distances, while only the disk
model shows the counter-bridge prominently extending to distances of $\sim$80 kpc.}
\label{fig_dbmodels}
\end{figure}

Figure \ref{fig_dbmodels} shows the distance--$\Delta\alpha$\footnote{Where $\Delta\alpha$ is the offset (in true angle)
in right ascension from the SMC center.} distribution of particles near the SMC in the
DB12 disk (a) and DB12 spheroid1 (b) models.  Both models show extensions to the east forming the well-known structure
of the Magellanic Bridge, prominently seen in \hi and young stars ($\lesssim$200 Myr) forming in the gas.  The model predictions
of a significant number of particles towards the east and at closer distances, with very few to the
west at those same distances, is quite similar, qualititatively, to what is seen in our stellar data.  Therefore, we find that
the newly-found stellar component to the eastern side of the SMC may be an intermediate-age/old ($\sim$1-12 Gyr) stellar component
of the tidally-stripped Magellanic Bridge (further discussed in next section).

\section{Discussion}
\label{sec:discussion}

We detect a large line-of-sight depth ($\sim$23 kpc) in our four eastern fields
covering at least 135\dgr in position angle.  The western fields have a much
shallower depth of $\sim$10 kpc with a quite sudden increase in depth between
PA=341\dgr and 26\degr.  Three of the eastern fields (PA=71\degr, 116\degr, and 161\degr)
show a distance bimodality with the farther component having $d$$\sim$67 kpc, similar to
the distance of the main body of the SMC and the western fields, and the closer component
at $d$$\sim$55 kpc, between the SMC and LMC distance.  The fourth eastern field (PA=26\degr)
has a large line-of-sight depth and is potentially consistent with the trend seen in the other
three eatern fields.
This is the first clear evidence of a distance bimodality
in the eastern SMC and a newly identified structure (the component at $\sim$55 kpc) which
we call the SMC ``eastern stellar structure''.

In Section \ref{sec:models} we compared our data to Magellanic Clouds interaction models
\citep[][see Fig.\ \ref{fig_models}]{Connors06,DB12,Besla12}.  Overall the models do not match the
data very well.  The Connors, DB12 spheroid1 and B12 models do not show the large depth in the
eastern fields that is seen in the data.  In contrast, the DB12 disk model shows a large
depth in some western and northwestern fields and a distance bimodality in northern and northeastern
fields.  The two components are from the main SMC body and the ``counter-bridge'',
which is a tidal stream at large distances ($\sim$80 kpc) and behind the SMC.  While the
model does have ``a'' bimodality, the distances do not match the data.  In contrast, all of the observed
fields show a component at $d$$\sim$66 kpc and the eastern fields have an extra component {\em in front}
of the SMC at $d$$\sim$55 kpc, with almost no stars beyond 70 kpc.  Therefore, there is no sign of the
counter-bridge in our stellar sample.  However, the DB12 disk model does show
a small number of particles between $\sim$50--60 kpc in the northeastern and eastern fields
that are not seen in the other fields.  In fact, when all particles with $d$$<$55 kpc are selected
they cover a wide region in the eastern SMC spanning PA$\approx$321--144\dgr (range of 183\degr).  The DB12 spheroid1
model shows a similar pattern of particles at this distance but at lower density and spanning
a smaller position angle range of 121\dgr (PA$\approx$65--186\degr).  It is possible that this is the
feature that we are detecting, although at a higher density than predicted by the models.  At larger radii
this model feature extends to even smaller distances and towards the LMC (see Fig.\ \ref{fig_dbmodels}).  For the
DB12 disk model, which is supposed to represent the gaseous component of the SMC, this arm should represent
the well-known \hi Magellanic Bridge.  If our new component at $R$=4\dgr and $d$$\sim$55 kpc is related
to this feature then we very well might be seeing, for the first time, a stellar component of the Magellanic Bridge.

\begin{figure}[t]
\begin{center}
\includegraphics[angle=0,scale=0.57]{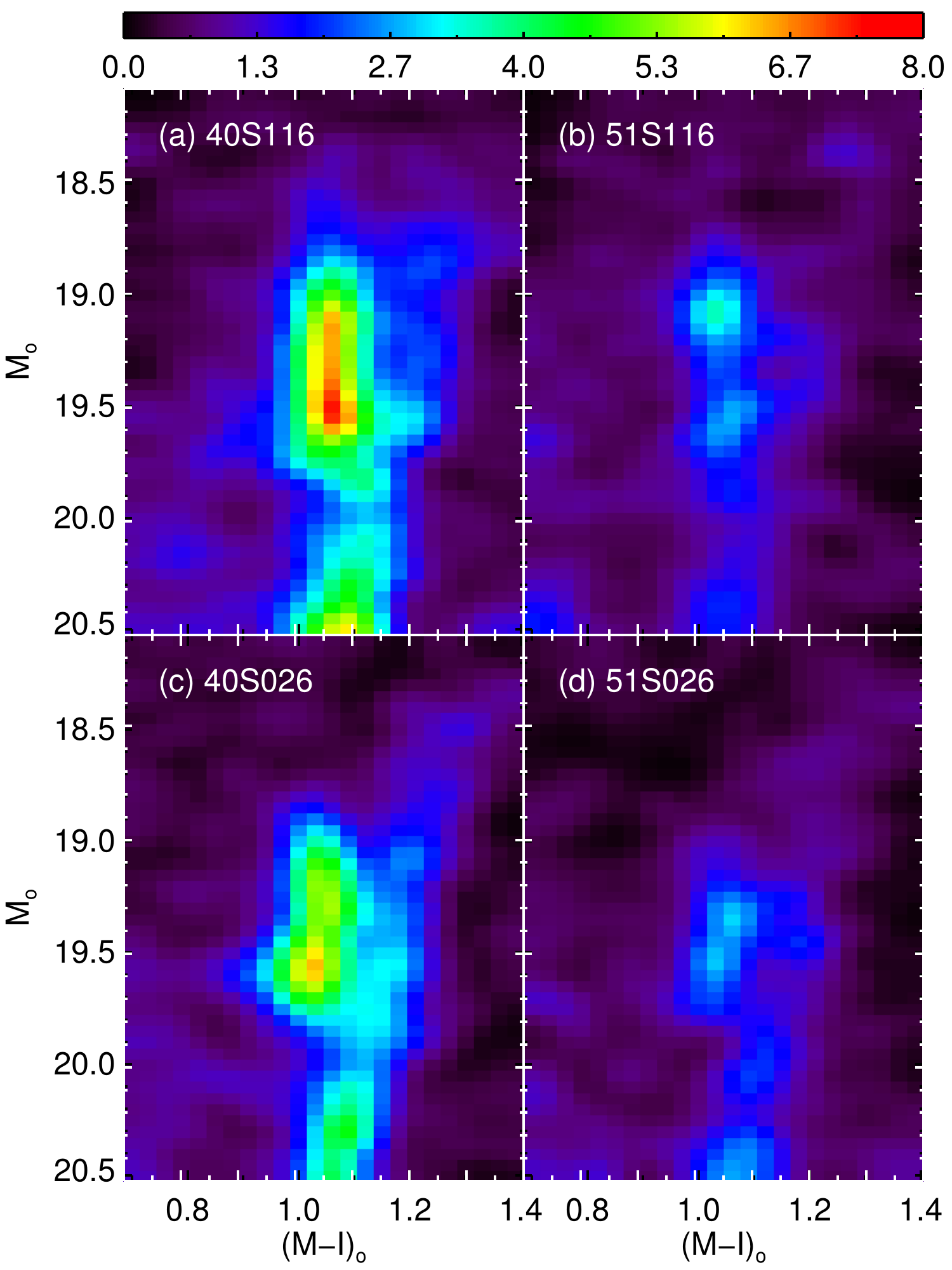}
\end{center}
\caption{Hess diagrams showing the dependence of the density of the two distance components (at $\sim$55 kpc and
$\sim$67 kpc) with SMC radius for two position angles. {\em (a)} 40S116, {\em (b)} 51S116, {\em (c)} 40S026, and
{\em (d)} 51S026.  At PA=116\dgr the density of both components drops with radius but the distant component
more quickly than the closer component.  At PA=26\dgr the density of both components again drops with radius but
this time the closer component drops more rapidly (almost vanishing) than the distant component.}
\label{fig_cmdpanels45}
\end{figure}

Even though the proximity of the new eastern structure to the center of the SMC and
its large extent argues for an SMC origin, we must consider other possibilities.
Could this be a stream of the LMC, a satellite of the SMC, or something else entirely (e.g., MW halo substructure)?
The new structure is likely not related to the LMC because the RC color is too blue.  The LMC is more
metal-rich ([Fe/H]$\sim$$-$0.4) than the SMC ([Fe/H]$\sim$$-$1.0; Pagel \& Tautvaisiene 1998) and this difference would be 
evident in the mean RC color, which is metallicity-dependent \citep{Girardi98}. For a 3.2 Gyr population (log(age)=9.5),
the Padova isochrones \citep{Girardi02} give a mean RC color of $M$-$T_2$$\approx$1.05 for [Fe/H]=$-1.0$ and
$M$-$T_2$$\approx$1.22 for [Fe/H]=$-0.40$ with a difference of $\sim$0.17 mag.  On the other hand, the observed mean
RC colors in our MAPS LMC and SMC dereddened CMDs show a difference of $\sim$0.04--0.05 mag, a smaller difference
than from the theoretical isochrones, likely because we are sampling the more metal-poor peripheries of both objects.
However, even a difference of $\sim$0.05 mag between the two components would be visible in the CMDs studied here.
While there are some small changes in mean RC color with magnitude they are not more than $\sim$0.01--0.02 mag
(Fig.\ \ref{fig_rcpanels}).  Furthermore, for an LMC-origin, the density of the new structure should increase
towards the LMC.  However, we have two eastern fields at $R$=5.1\dgr from the SMC (closer to the LMC than the
four eastern fields analyzed here) that have extended RCs but at {\em lower} densities than in the $R$=4.0\dgr
fields (Fig.\ \ref{fig_cmdpanels45}).  This indicates that the new structure decreases in density from the SMC
center, but not from the LMC center.  The new structure is, therefore, unlikely to be related to the LMC.

The large extent in position angle of this structure (corresponding to $\sim$9 kpc), and its fairly uniform density across
that distance, makes it unlikely
to be a completely new satellite galaxy.  A stream of a satellite galaxy of the SMC could span such a large
region of the sky, but the density of the stream would have to rival that of the SMC itself in those regions
(and exceed it in some places), which would imply a truly massive satellite and a core that should have been previously detected.

A new MW halo substructure is also unlikely because it would need to have nearly the identical metallicity, position in
the sky, and distance (closer by $\sim$10 kpc) as the SMC (producing a nearly identical distribution in the CMD) and,
additionally, have a density fall-off with SMC radius.  Therefore, we conclude that the most likely explanation is that the
new component at $d$$\sim$55 kpc is a stream of the SMC itself.  The new component's location (to the east) and distance
roughly match that expected for a tidally stripped stream of stars from the \citet{DB12} simulations that were
``loosened'' in the last close encounter of the MCs $\sim$200 Myr ago.

While it is more difficult to use RC stars to study the depth of the inner SMC (where there are many young
stars) because of age effects, nevertheless such an extended RC as seen in the eastern SMC might be detectable.
The RC luminosity function in the inner SMC ($R$$\lesssim$2\degr), using MCPS \citep{ZH02} and OGLE-III \citep{ogle3} data,
looks much more like those in our western fields than our eastern fields, and there is little variation
in the shape of the RC with position angle and radius.
However, there are three reasons why we are not likely to detect much structure
in the density function with distance near the center of the SMC:
(1) Near the center the density is very centrally-peaked (in distance) making it difficult to detect a lower-density structure
at a non-systemic distance; at larger radii the density distribution
is much less centrally-peaked and it becomes easier to detect deviations. (2) It is more difficult to strip stars from the
center (because they are more tightly bound) than from the periphery.
(3) For stripped stars, deviations in distance from the systemic value should grow with radius (see Fig.\ \ref{fig_dbmodels}).
Therefore, it is not too surprising that the RC shape in the central region of the SMC looks quite regular and any deviations
become visible only at larger radius.
We note that the central SMC RR Lyrae also show little spatial pattern or variations and have a depth of only
$\sim$8 kpc, similar to what we observe in our western fields \citep{Haschke12}.

In the near future several wide-field photometric surveys will be able to provide the data needed to study the 3D structure
of the SMC periphery in great detail.  OGLE-IV \citep{ogle4} will provide high-quality time-series photometry with which
RR Lyrae and Cepheids can be identified and accurate distances measured, as was done by \citet{Haschke12} with OGLE-III.  OGLE-IV,
SkyMapper \citep{Keller12} and DES \citep{Abbott12}, as well as other DECam programs, will provide photometry to
well-below the SMC horizontal branch over a large area of the MCs and with which RC stars can be exploited to
study the 3D structure of the SMC.

We plan to study the radial velocities and metallicities of stars in the SMC periphery, especially in the east,
to help understand any kinematical or chemical differences that may exist between the two components
\citep[as previously seen by][]{Hatz93}
and that might shed more light on the origin of the newly found stellar structure.

Finally, we note that the existence of a stellar structure {\em in front} of the main SMC stellar population could have
important implications for microlensing surveys, in that this will increase the self-lensing of SMC stars
\citep{Besla13,Calchi13}.  We recommend that this newly identified structure be taken into account in the analysis
of microlensing surveys probing the eastern periphery of the SMC (e.g., OGLE-IV).

\section{Summary}
\label{sec:summary}

We use high-quality CTIO-4m+MOSAIC photometry in eight fields at $R$=4\dgr in the SMC to study the outer galaxy's
line-of-sight distribution.  Many of the fields show very extended red clump luminosity distributions, as previously
seen by \citet{HH89} and \citet{GH91}.  We show that the extended red clump luminosity distributions cannot be
accounted for by age effects because the main-sequence counterparts of very young populations ($\lesssim$1 Gyr)
are not observed.  Our main results and conclusions are:
\begin{enumerate}
\item The four eastern fields show very large line-of-sight depths ($\sim$23 kpc) over $\sim$135\dgr of position angle.
\item Three eastern fields show a strong distance bimodality with one component at $\sim$67 kpc (near the mean SMC distance)
and a second component at $\sim$55 kpc.  The fourth eastern field (40S026) has a broadened distance distribution, and
is potentially consistent with the trend seen in the other three eastern fields but at slightly larger distances.
\item The newly-found stellar component in the east at $\sim$55 kpc is qualitatively consistent with the \citet{DB12}
model distribution of particles in the tidally-stripped Magellanic Bridge, previously only detected in \hi.  We
conclude that this new component is likely an intermediate-age/old ($\sim$1-12 Gyr) stellar component of the Magellanic
Bridge and call it the SMC ``eastern stellar structure''.
\end{enumerate}

A tidally-stripped stellar component of the Magellanic Bridge is consistent with the discovery of accreted SMC stars
in the LMC by \citet{Olsen11} and the claim by \citet{Besla13} of a tidal origin for the microlensing events reported
towards the LMC.  In the future, we plan to follow-up our discovery using spectroscopy in SMC fields to
compare the stellar velocities to those predicted by the models for the Magellanic Bridge.

We find that even though there are some similarities between our data presented here and models from the literature,
the differences are much more apparent and it is clear that more work is needed on the simulations to match the
SMC stellar distribution.  It might be that the stellar components of the SMC (disk or halo) are initially more extended
than the simulations have so far considered.

\acknowledgements

We dedicate this paper to Robert T. Rood who did pioneering work on
horizontal branch stars and found the extended SMC RC very fascinating.
We thank J.D. Diaz, Gurtina Besla and Mario Mateo for useful discussions,
and Diaz, Besla and Connors for sharing their models with us so we could
compare them to our data.  We also thank Sebastian Hidalgo for running
IAC-star population synthesis models for us.
We thank the OGLE and MCPS projects for making their SMC photometric databases available
to the public, and Despina Hatzidimitriou for providing us with her photographic plate
photometric of the SMC periphery.
D.L.N. was supported by a Dean B.\ McLaughlin fellowship
at the University of Michigan.
E.F.B. acknowledges support from NSF grant AST 1008342.
We acknowledge funding from NSF grants AST-0307851 and AST-0807945,
and NASA/JPL contract 1228235.
R.R.M.~acknowledges support from CONICYT through project BASAL PFB-06
and from the FONDECYT project N$^{\circ}1120013$.

{\em Facilities:} CTIO (MOSAIC II).

%  THE BIBLIOGRAPHY

\end{document}